\documentclass[aps,pra,twocolumn,showpacs]{revtex4-1}

\usepackage{graphicx,epstopdf}
\usepackage{amsmath}
\usepackage{amssymb}
\usepackage{braket}
\usepackage{color}
\usepackage{float}
\usepackage{epstopdf}

\usepackage[colorlinks=true, citecolor=blue, urlcolor=blue ]{hyperref}
\input{epsf}
\begin{document}

%
%

\title{Enhancement of Quantum Correlation Length in Quenched Disordered Spin Chains}
\author{Debasis Sadhukhan, Sudipto Singha Roy, Debraj Rakshit, R. Prabhu, Aditi Sen(De), and Ujjwal Sen}
\affiliation{Harish-Chandra Research Institute, Chhatnag Road, Jhunsi, Allahabad 211 019, India}

\date{\today}

\begin{abstract}

Classical correlations of ground states typically decay exponentially and polynomially, respectively for gapped and gapless short-ranged quantum spin systems. In such systems, entanglement decays exponentially even at the quantum critical points. However, quantum discord, an information-theoretic quantum correlation measure, survives long lattice distances. We investigate the effects of quenched disorder on quantum correlation lengths of quenched averaged entanglement and quantum discord, in the anisotropic $XY$ and $XYZ$ spin glass and random field chains. We find that there is virtually neither reduction nor enhancement in entanglement length while quantum discord length increases significantly with the introduction of the quenched disorder.
\end{abstract}


\maketitle

\section{Introduction}
\label{sec_introduction}

In the last twenty years or so, it has undoubtedly been established that entanglement \cite{Horodecki} plays a significant role in efficient quantum communication protocols~\cite{AditiUjjwal}, which include quantum dense coding \cite{densecode}, quantum teleportation \cite{teleportation} and quantum key distribution \cite{BB84}, and in 
quantum computational tasks like the one-way quantum computer~\cite{briegel}.
 Generation of entanglement in composite systems, requires interaction between its subsystems and hence strongly interacting systems, such as quantum spin chains, form natural resources of entanglement. In the last decade, it has also been realized that entanglement can be used to detect low-temperature phenomena in many-body systems \cite{manybody-entanglement}. Spin models have further been identified as ideal setups  for transferring quantum states \cite{bose} as well as for realizing quantum gates, essential parts of quantum circuits and quantum simulators \cite{quant-compu-1, briegel}. Moreover, spin models can be prepared in laboratories via e.g. ultracold atoms in optical lattices \cite{cold-atom}, trapped ions \cite{ions}, and nuclear magnetic resonance systems \cite{NMR} in a controlled way, and also can be found in materials \cite{heisenberg-material, order-from-disorder-1}.

                                                                                                                                                                                                                                                                                                                                                                                                                                                                                                                                                                                                                                                                                                                                                                                                                                                                                                                                                                                                                  Distribution of entanglement among various parts of a many-body system is important for transmitting information from one party to another within  communication
computing networks  \cite{bose,network, briegel}.                                                                                                                                                                                                                                                                                                                                                                                                                                                                                                                                                                                                                                                                                                                                                                                                                                                                                                            The efforts to understand the sustainability of correlations in many body systems as a function of lattice distance happens to be crucial, since the low-temperature scaling of correlations are expected to remain universal, irrespective of the microscopic details of various materials governed by the same class of underlying models. In case of gapped systems with short-ranged interactions, using the {{Lieb-Robinson bounds}} \cite{lieb-robin}, it can be shown that two-site classical correlation functions, $\langle { \cal {O}}_i {\cal{O}}_j\rangle-\langle {\cal{O}}_i\rangle \langle{\cal{ O}}_j \rangle$,   decay exponentially with increasing  lattice distance,  known as {{clustering of correlations}} \cite{eisert_tensor, hasting} except when the systems are critical, when they decay polynomially,  with ${\cal{O}}_i$ and ${\cal{O}}_j$ being the observables at sites $i$ and $j$ respectively. However, it has been reported that in many cases, quantum correlations, as qualified by entanglement,  decays much more rapidly in comparison to classical correlators, both near and far from the quantum critical points.  For example, it was shown that in the quantum $XY$ model with transverse magnetic field, entanglement can survive only up to the next nearest neighbor \cite{phase-transition-ent}, while in the Heisenberg model, it becomes vanishingly small already after the nearest neighbor \cite{hisen_length}.

                                                                                                                                                                                                                                                                                                                                                                                                                                                                                                                                                                                                                                                                                                                                                                                                                                                                                                                                                                                                            In recent times, information-theoretic quantum correlation measures, like quantum discord \cite{ref-discord} and quantum work deficit \cite{ref-wd}, were introduced which promise to estimate quantum correlations {\it beyond} entanglement \cite{negative-eff-defect}. For example,  it has been shown that there exist instances in which quantum discord is capable of detecting quantum phase transition  while pairwise entanglement fails, especially at finite temperature \cite{phase-transition-discord1,phase-transition-discord}. Hence, it is  natural to investigate the scaling of these two-site information-theoretic quantum correlation measures in these systems, with increasing distance between the sites. In this respect, it is already known that unlike entanglement, quantum discord can survive over a longer distance \cite{sarandy-DL,amico-DL} in the ordered $XY$ spin chain.

Establishing finite correlations between distant parties is undoubtedly important to implement quantum information processing tasks in many body systems. In this paper, we ask the following question: {\em Is it possible to enhance quantum correlation lengths, namely entanglement and quantum discord lengths, significantly, by introducing defects in  quantum spin systems?} We find that the answer is in the affirmative. Disorder occurs unavoidably in real materials and can now also  be engineered artificially in, e.g. cold atom experiments \cite{disorder-expt}. Moreover, it was shown that there exists some models, in which disorder plays a constructive role by enhancing physical properties like magnetization, classical correlators, and quantum correlations \cite{order-from-disorder-1, order-from-disorder-2, order-from-disorder-3}. Such phenomena, known as ``order-from-disorder'' or ``disorder-induced order", run certainly in contrary to the naive belief that impurity in the systems can have only a debilitating effect.

In this work, we investigate quantum correlation lengths for the ground states of one-dimensional quenched disordered anisotropic $XY$  and $XYZ$ spin chains. Our results show that although the entanglement lengths cannot be improved  significantly in these quenched disordered systems, the lengths of information-theoretic quantum correlation measures, can  substantially be enhanced in such systems. Such disorder-induced advantages can be seen in quantum $XY$ spin glass and random field $XY$ models, where the investigations are carried out with the help of Jordan-Wigner transformations, and in the $XYZ$ spin glass systems, which are handled by using density matrix renormalization group (DMRG) techniques.

The rest of the paper is arranged as follows. Section~\ref{sec:models} introduces the models under study, and discusses the methods involved in solving them. Section~\ref{sec:measures} reviews various quantum correlation measures used in this work and introduces  quantum correlation lengths. The results for the disordered $XY$ and $XYZ$ models are discussed in Secs.~\ref{sec:XY} and ~\ref{sec:XYZ} respectively. Finally, we summarize our results in Sec.~\ref{sec:conclusion}.

\section{Models and Methodology }
\label{sec:models}

 In this section, we briefly describe the models that we use in this paper. Among the models described, note that the ordered $XY$ spin model is exactly solvable and the analytic technique 
 can be used to handle the corresponding disordered model, while the $XYZ$ spin model with and without disorder cannot be solved analytically. 
 We also present a brief decription of quenched averaging. 
 
 \subsection{Quenched disorder and averaging}
The disorder in the system parameters are taken to  be ``quenched".
That is, we assume that the time scale over which the system dynamics of interest takes place is much smaller compared to the time scale over which there is a change in the particular set of parameters governing the disorder in the system. In order to calculate the quenched averaged value of a physical quantity, we need to perform the averaging over the probability distribution of several realizations, each of which corresponds to a fixed configuration of the system, after calculating the value of the physical quantity for the fixed configurations.

\subsection{Quantum $XY$ spin chain: Ordered and Disordered Models}
\label{subsec:model_xy}
The general Hamiltonian for the quantum $XY$ spin chain with nearest-neighbor interactions
in an external magnetic field is given by
\begin{align} 
H =   \kappa \Big[ \sum_{i = 1}^{N} \frac {J_i}{4} \Big((1+\gamma) \sigma_i^x \sigma_{i+1}^{x} + (1 - \gamma ) \sigma_i^y \sigma_{i+1}^{y}\Big) \nonumber \\
-  \sum_{i=1}^N \frac {h_i}{2} \sigma_i^z \Big],
\label{eqn: XY}
\end{align}
where $\kappa J_i$  are the coupling constants, $\kappa h_i$ is the magnetic field strength at the $i^\text{th}$ site,  and $\gamma$ is the anisotropy constant.   The constant $\kappa$ has the units of energy, while $J_i,h_i,$ and $\gamma$ are dimensionless. Here, $\sigma^j,$ for $j=x,y,z,$ correspond to the Pauli spin matrices. Moreover,  we  assume the periodic  boundary condition, i.e., $\vec{\sigma}_{N+1}=\vec{\sigma}_{1}$.

 {\bf{Case 1: Quantum $XY$ spin glass::}} In this case, the coupling strengths $J_i$ are randomly chosen from independently and identically distributed ($i.i.d.$) Gaussian distribution with mean $\langle J \rangle$ and unit standard deviation. However, the field is kept  uniform throughout the lattice, i.e., $h_i=h$ for $i=1,\cdots,N$. 
 
  {\bf{Case 2: Random field quantum $XY$ spin chain::}} The model is now with uniform coupling, i.e., $J_i=J$ but the $h_i$ are $i.i.d.$ Gaussian random variables with mean $\langle h \rangle$ and unit standard deviation.
  
 {\bf{Case 3: The ordered quantum $XY$ spin chain::}} 
 The model is with site-independent 
  coupling constants as well as the magnetic field strengths, i.e., $J_i = J$ and $h_i = h$ for $i=1,\cdots,N$. 

The ordered quantum XY spin chain is exactly solvable via successive applications of the Jordan-Wigner, the Fourier, and the Bogoliubov transformations~\cite{LSM,  barouch1, barouch2}. The corresponding disordered systems can also be handled upto relatively large system sizes by using the same transformations. The two-site reduced density matrix, $\rho_{ij}$, of the ground state can  be easily constructed from the one and two-point correlation functions as
\begin{align}
\rho_{ij} = \frac{1}{4}\Big[I\otimes I+&m^z_i(\sigma^z\otimes I) + m^z_{j}(I \otimes \sigma^z)  \nonumber\\
 &+ \sum_{\alpha=x,y,z}T_{ij}^{\alpha\alpha}(\sigma^{\alpha}\otimes\sigma^{\alpha})\Big]. \label{eqn: rho12}
\end{align}
Note that $m_i^x = m_i^y =0$ and all the off-diagonal correlations vanish.

\subsection{Quantum $XYZ$ spin glass}

Here, our interest lies in the quenched averaged correlation lengths in the $XYZ$ spin chain. The Hamiltonian of the quantum $XYZ$ spin glass model is given by 
 \begin{eqnarray} \label{eqn: XYZ}
 H = \kappa\Big[\sum_{i=1}^{N-1}\Big[ \frac {J_{i}}{4}\big[(1+\gamma)\sigma_i^x\sigma_{i+1}^x+(1-\gamma)\sigma_i^y\sigma_{i+1}^y\big]&+ \nonumber \\
\frac {\Delta}{4} \sigma_i^z\sigma_{i+1}^z\Big] 
-\frac{h}{2}\sum_i \sigma_i^z \Big],&
\end{eqnarray}
where $\kappa \Delta$ is the nearest-neighbor coupling strength for the $zz$- interaction, which is independent of the site. The rest of the parameters are the same as discussed above in the context of the $XY$ spin glass. The corresponding ordered Hamiltonian can be obtained from Eq.~(\ref{eqn: XYZ}) by simply setting $J_i=J$ for $i=1,\dots,N$.

Unlike the quantum XY disordered chain, for which the ground state of a considerably large number of spins can be obtained by using the Jordan-Wigner transformation, one needs to resort to numerical techniques for the XYZ model with random coupling strengths. In order to investigate the  ground state for the system characterized by the Hamiltonian in Eq. (\ref{eqn: XYZ}), we employ the well-established numerical technique called the DMRG method \cite{dmrg}. After performing the standard infinite size DMRG, several finite size DMRG sweeps are also carried out in order to increase the accuracy of the calculations for the inhomogeneous chain.

\section{Quantum correlation measures}
\label{sec:measures} 
The presence of quantum correlations  between subsystems of a composite system helps in realizing many quantum information protocols. In order to explore these protocols, it is necessary to quantify the quantum correlations involved. In this work, we have mainly used two quantum correlations measures, namely, concurrence and quantum discord. They belong to two different paradigms of quantum correlation -- while the first corresponds to the entanglement-separability paradigm, the other is in the information-theoretic one.  In the following subsections, we will briefly  introduce both the quantum correlation measures considered here. We will then provide a short introduction of the concept of quantum correlation lemgth. 

\subsection{Concurrence}
\label{subsec:conc}
Concurrence \cite{wooters} quantifies the amount of entanglement present in an arbitrary two-qubit state. Given a two-qubit density matrix, $\rho_{AB}$, the concurrence is defined as 
\begin{equation} 
 C(\rho_{AB})=\max \{0, \lambda_1 - \lambda_2 - \lambda_3 - \lambda_4\},
\end{equation}
where $\lambda_i$'s are the eigenvalues of the Hermitian  matrix  $ R=\sqrt{\sqrt{\rho}~\widetilde{\rho}~\sqrt{\rho}}$ and satisfy the order $\lambda_1\geq \lambda_2\geq \lambda_3\geq \lambda_4$. Here $ \widetilde{\rho}=(\sigma_y \otimes \sigma_y) \rho^\ast (\sigma_y \otimes \sigma_y)$ with  $\rho^*$ being the complex conjugate of $\rho$ in the computational basis. 

\subsection{Quantum discord}
\label{subsec:QD}
In classical information theory, the amount of ignorance about a probability distribution, ${p_i}$, is quantified by the Shannon entropy, defined as $H(\{p_i\})=-\sum_i  p_i \log_2 p_i $. The mutual information between two classical random variables $i$ and $j$, having the marginal distributions $\{p_i\}$ and $\{p_j\}$, can be defined in two equivalent ways as
\begin{align}
\begin{split}
{\cal I}(\{p_{ij}\}) &= H(\{p_i\})+H(\{p_j\})-H(\{p_{ij}\})\\ \label{eq:mutual1}
                              &=H(\{p_i\})-H(\{p_{i|j}\}),
\end{split}
\end{align} 
where $\{p_{ij}\}$ and $\{p_{i|j}\}$ correspond to  the joint probability distribution of the variables $i$ and $j$, and the conditional probability distribution respectively. In case of quantum systems, the quantity
  \begin{equation}
 {\cal I}(\rho_{AB})=S(\rho_A)+S(\rho_B)-S(\rho_{AB}), \label{eqn:dis}
  \end{equation}  
for a two-party quantum system $\rho_{AB}$, can be argued to quantify the total correlation present in the system  where $S(\sigma)=-\text{tr}[\sigma \log_2 \sigma]$ and $\rho_i$, $i=A,B$ are the local density matrices of $\rho_{AB}$ \cite{mutual1, mutual2, mutual3}. This quantity can be interpreted as the quantized version of the first expression of the classical mutual information in Eq.~(\ref{eq:mutual1}). To quantize the other expression, we consider that a  measurement is performed on one party, say $B$, using a complete set of rank-one projectors, $\{B_i\}$, satisfying the relations $B_i B_j =\delta_{ij} B_i$ and $\sum_i  B_i=I_B$. The post-measurement ensemble is given by $\{p_i, \rho^i_{AB}\}$ with $\rho^i_{AB}=(I_A\otimes B_i)~ \rho~  (I_A\otimes B_i)$ and $p_i=\text{tr}((I_A\otimes B_i) ~\rho_{AB}~  (I_A\otimes B_i))$, where $I_A$ is the identity operator on the Hilbert space of the system with observer $A$. The corresponding quantum conditional entropy is given by
   \begin{equation}
    S(\rho_{A|B})=\min_{\{B_i\}}\sum_i p_i  S(\rho_{A|i})\label{eqn: min},
\end{equation}
where $\rho_{A|i}=\text{tr}_B \rho^i_{AB}$. The quantum version of the second expression in 
Eq.~(\ref{eq:mutual1}) of classical mutual information then reads 
   \begin{equation}
   {\cal J}(\rho_{AB})= S(\rho_{A})-S(\rho_{A|B}), \label{eqn: cld}
   \end{equation} 
   which turns out to be inequivalent to the expression in  Eq.(\ref{eqn:dis}), and can be argued as a measure of classical correlation of the state $\rho_{AB}$. Quantum discord \cite{ref-discord} is defined as the difference between these two inequivalent quantities, and is given by
  \begin{equation}
   {\cal D}(\rho_{AB})={\cal I}({\rho_{AB}})-{\cal J}(\rho_{AB}).\label{eqn:discord}
   \end{equation}

In all the cases considered in this paper, the bipartite states are $X$ states \cite{x-states} with $|T^{xx}_{ij}|\geq |T^{yy}_{ij}|$, for which an analytical form of quantum discord is available \cite{Ali-et-all,Wang,Chen,Huang-Disc-num}. We have also checked the claim numerically. When the measurement is performed by the first party, i.e. on the $i^\text{th}$ spin of a two-party state $\rho_{ij}$, the classical correlation is given by $ {\cal J}(\rho_{i,j}) = H_2\left(\frac{1+m_i^z}{2}\right) - H_2\left(\frac{1+p}{2}\right) $, with $p=\sqrt{(m_i^z)^2 + (T^{xx}_{i,j})^2}$ and $H_2(x)=-x\log_2x-(1-x)\log_2(1-x)$ being the binary entropy.

\subsection{Quantum correlation length}
\label{sec:QCorrLength}
As mentioned earlier, our primary aim is to investigate quantum correlation length in quenched disordered $XY$ and $XYZ$ spin models. 
For arbitrary ordered spin systems, let us first define the entanglement length, in particular, concurrence length. If the concurrence, $C_{i,j}$ between the $i^\text{th}$ and $j^\text{th}$ spins behaves as  
\begin{equation}
C_{i,j} \sim e^{-\frac{|i-j|}{\xi_C}}, \label{eqn:EL_C}
\end{equation}
we refer to $\xi_C$ as the concurrence length. For quenched disordered systems, the concurrence length for the quenched averaged two-site concurrences, if the latter behaves as in Eq.~(\ref{eqn:EL_C}), will be denoted by $\xi_{\langle C \rangle}$. In this paper, entanglement length will mean concurrence length. 

 Similar to the concurrence length, one can define discord length, $\xi_D$, for the ordered models, if the quantum discord, $D_{i,j}$, between the $i^\text{th}$ and the $j^\text{th}$ sites behaves as  
 \begin{align}
 D_{i,j} \sim a+ b~ e^{-\frac{|i-j|}{\xi_D}},\label{eqn_EL_D}
\end{align}
where  $a$, $b$ are parameters that are determined by the model under consideration. 
And similarly as above, the discord length for quenched disordered spin models can be defined, and is denoted by $\xi_{\langle D \rangle}$.  
A note on the units used here and henceforth in the paper. The entanglement and discord length are in the units of lattice distance. The parameters $a$ and $b$ in Eq.~(\ref{eqn_EL_D}) and in similar equations are in bits, just as for quantum discord. We have kept silent a multiplicative constant on the right hand side of Eq.~(\ref{eqn:EL_C}) that has the unit of ebits, just as the concurrences.  

Note that this is different from the concept of {\it localizable entanglement}, the amount of entanglement that can be concentrated between two parties  through measurements performed on the rest of the parties \cite{LE} (see also \cite{dorit,popsescu_rol,asd_sen_marek}).

\section{Entanglement and Discord Lengths in anisotropic $XY$ Model} 
\label{sec:XY}

The ordered quantum $XY$ model is exactly solvable for finite as well as infinite spin chains~\cite{LSM,barouch1,barouch2}. In the thermodynamic limit of the system, closed form expressions for the magnetization \cite{barouch1} and the two-point correlation functions \cite{barouch2} are known, which are necessary to calculate the two-body density matrices between two arbitrary sites, and eventually to compute quantities like the concurrence and quantum discord. However, this is not the case for the disordered systems and one is ultimately restricted to study finite size systems, but techniques similar to those for the infinite system can help to compute quantum correlations for relatively large systems \cite{LSM,zanardi,group-disorder}.

 In order to make a comparison between the disordered and ordered systems, we fix the means of the distributions of the disordered parameters in the disordered systems to be identical to the corresponding parameters of the ordered system. In particular, the quenched averaged physical quantity, $Q_{av}(\langle a \rangle, \langle b \rangle, \dots)$, corresponds to the disordered system with disordered parameters $a, b, \dots$ having means $\langle a \rangle, \langle b \rangle, \dots$ respectively. The corresponding physical quantity for the ordered system is then $Q(\langle a \rangle, \langle b \rangle, \dots)$, where the values of the system parameters $a, b, \dots$ are kept constant at $\langle a \rangle, \langle b \rangle, \dots$ respectively.

  Note that in the case of the ordered $XY$ model, the ground state is multipartite entangled except at $J/h=1/\sqrt{1-\gamma^2}$ which is known as the factorization point \cite{facto1, facto2}. At this point, the ground state  is doubly degenerate and both the degenerate states are factorized as a tensor product of quantum states corresponding to all the individual spins. 
 Since we have taken the ground state to be a pure symmetric  state by taking equal superposition of both the degenerate ground states, the  two spin entanglement here vanishes at this point, while quantum discord may have a non-zero value. 
    
\begin{figure}[t]
\includegraphics[angle=0,width=4.2cm]{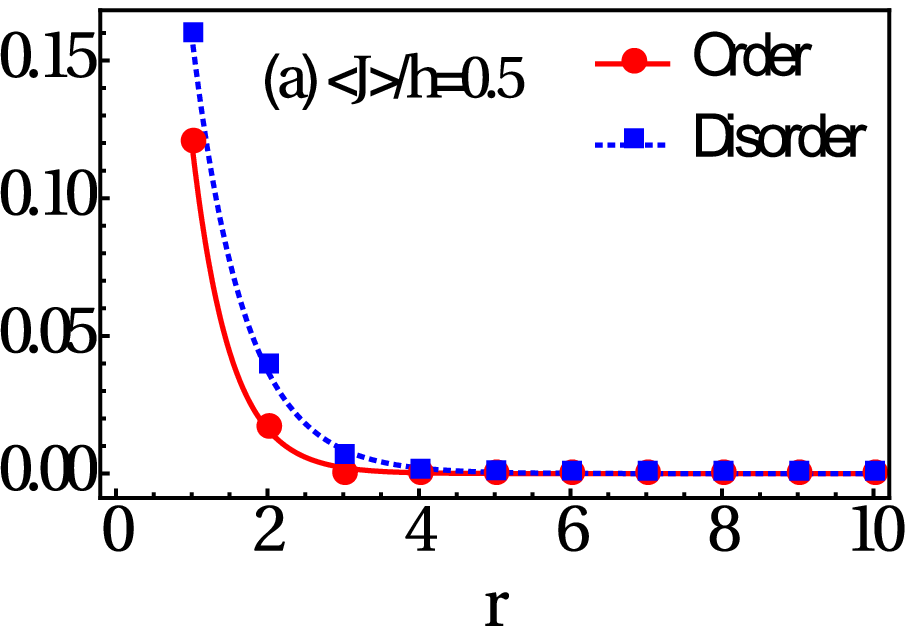} 
\includegraphics[angle=0,width=4.2cm]{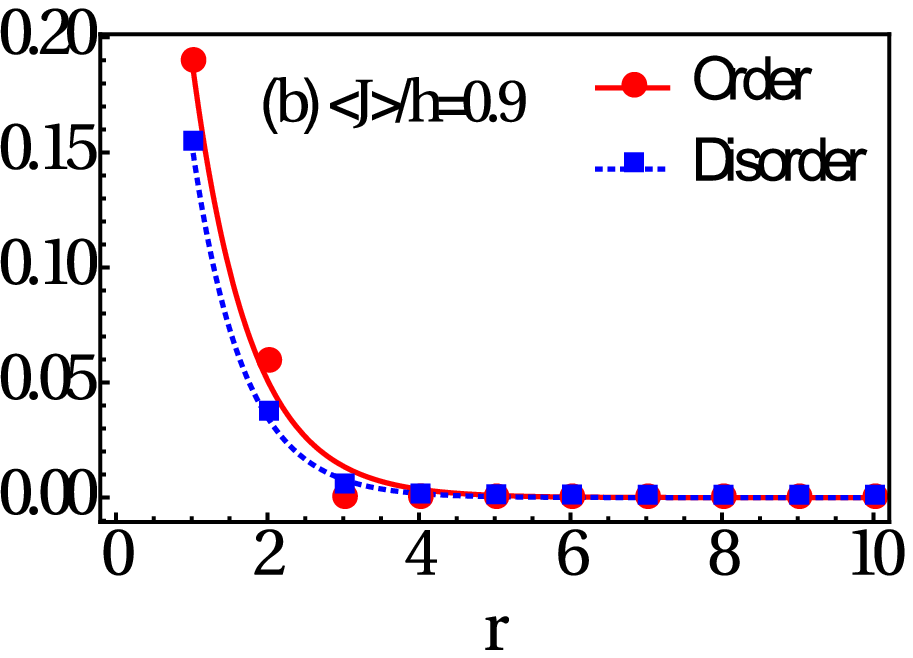} \\
\includegraphics[angle=0,width=4.2cm]{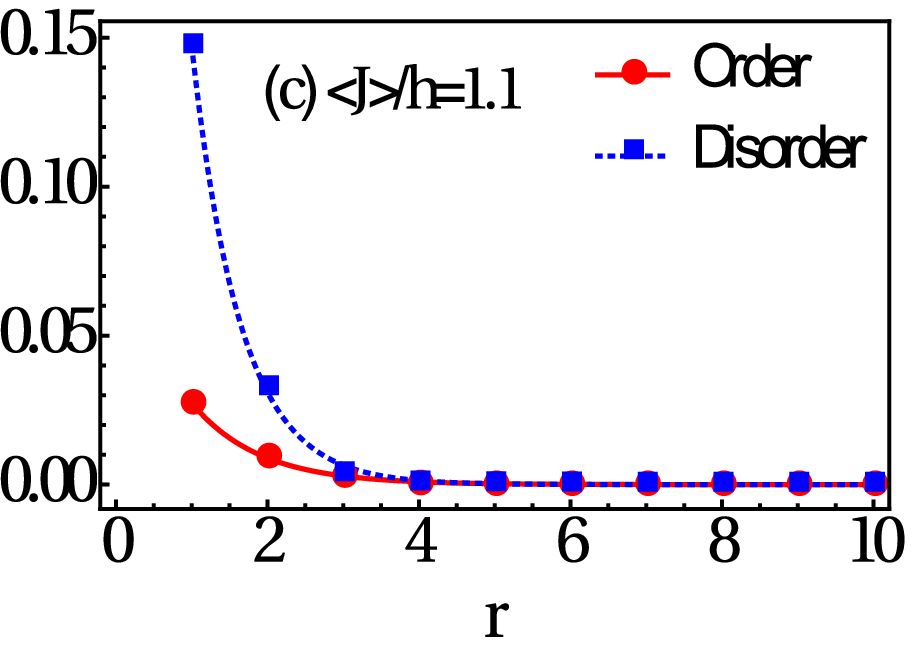} 
\includegraphics[angle=0,width=4.2cm]{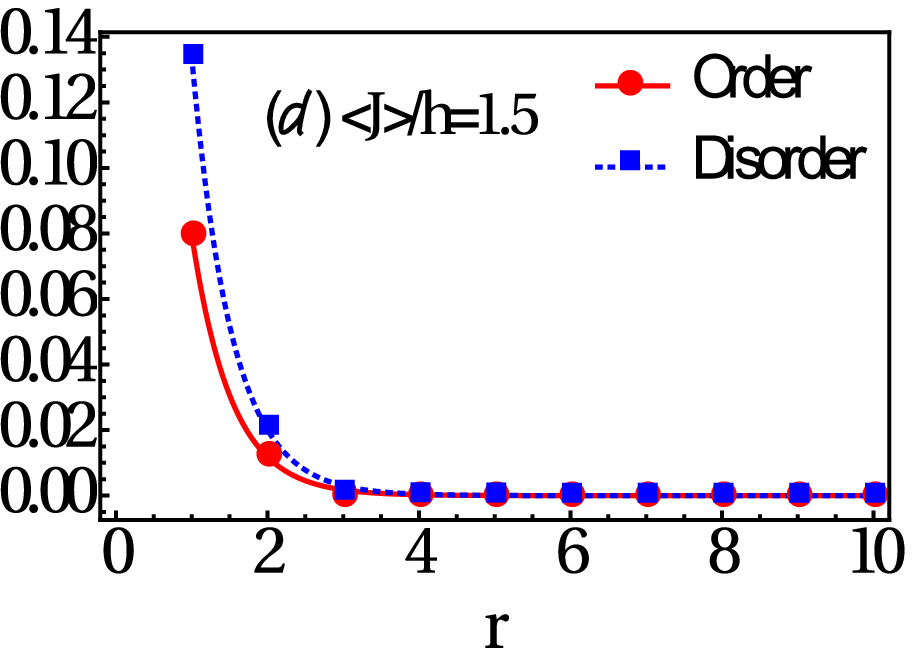} 
\caption{(Color online.) Entanglement length in quantum $XY$ spin glass vs. the ordered $XY$ model. In each panel, the dotted curve denotes the quenched averaged concurrence, $\langle C_{i,j} \rangle$, between the sites $i$ and $j$, plotted against the lattice distance $r=|i-j|$, for the quantum $XY$ spin glass for a certain value of $\langle J \rangle/h$. The corresponding curves for the ordered systems are also drawn in each panel. The vertical axis denote the concurrences, while the horizontal ones denote the lattice distances. Here, $N=50$ and $\gamma=0.5$. For the disordered case, the number of realizations of the random coupling is  taken to be $10^4$, for the quenched averaging. The lines are exponential fits and the obtained data at integer values of $r$. The vertical axis are measured in ebits, while the horizontal axis are in lattice length unit.     
}
\label{fig:decay_conc_xy_glass}
\end{figure}

\begin{figure}[t]
\includegraphics[angle=0,width=4.2cm]{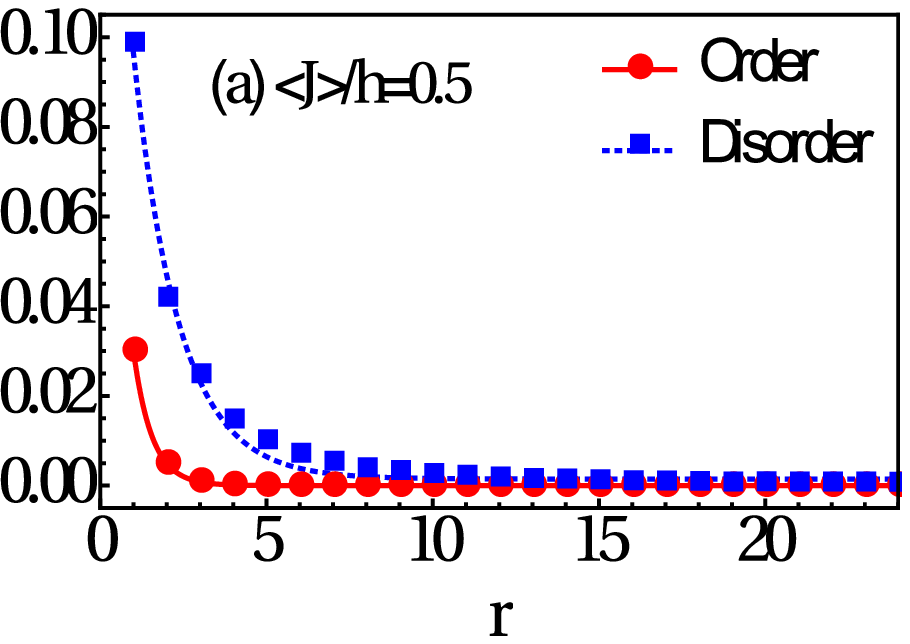} 
\includegraphics[angle=0,width=4.2cm]{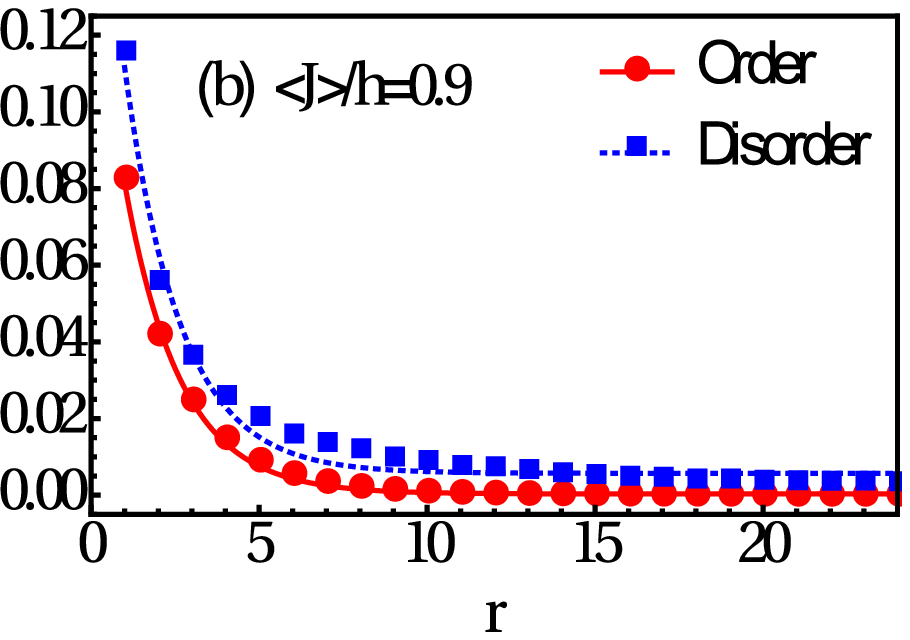} \\
\includegraphics[angle=0,width=4.2cm]{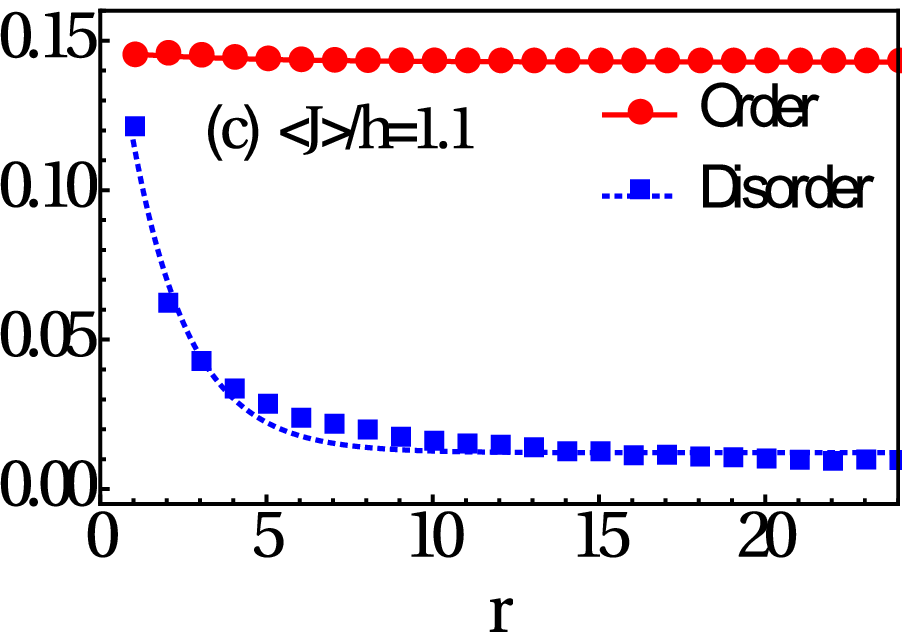} 
\includegraphics[angle=0,width=4.2cm]{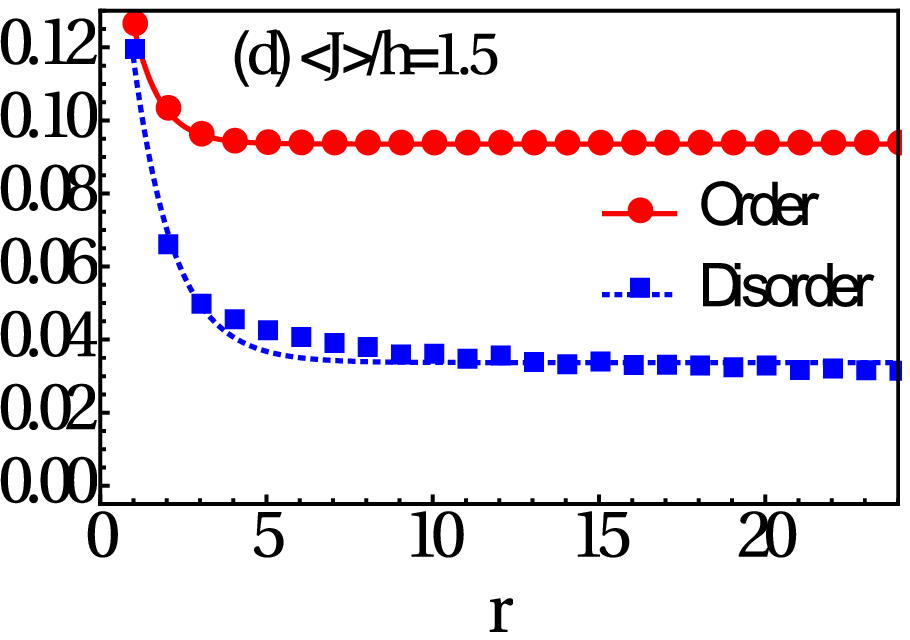} 
\caption{(Color online.) All considerations here are the same as in Fig.~\ref{fig:decay_QD_xy_glass}, except that quantum discord is considered instead of concurrence, and that the former is measured in bits.}

\label{fig:decay_QD_xy_glass}
\end{figure}

\subsection{Anisotropic quantum $XY$ spin glass: Entanglement length vs. Discord length}

Let us first consider the quantum $XY$ spin glass with $N$ sites. All considerations are for the pure symmetric ground state.

\subsubsection{Entanglement length}

We will now investigate the effects of impurities on entanglement length in the $XY$  spin model. 
To this end, we compare the two-site concurrence between the sites $i$ and $j$ of the ordered system, described by $H(J,h)$, with the quenched two-site concurrence between the same sites for the $XY$ spin glass system with $\langle J \rangle/h=J/h$, where $\langle J \rangle$ is the mean of the Gaussian distributed $J_i$ in the spin glass system (for all $i$). 
 In order to calculate the concurrence for the $XY$ spin glass model, we perform configurational averaging over $10^4$ random realizations.
 From  Fig.~\ref{fig:decay_conc_xy_glass}, we notice that  the concurrence for the ordered $XY$ model vanishes for $|i-j|\gtrsim 3$  while in  the disordered system, it goes to zero  for $|i-j|\gtrsim 4$,
 implying no significant enhancement or deterioration due to randomness.
   The lines in the panels in Fig. \ref{fig:decay_conc_xy_glass}, connecting the squares for the spin glass system and those connecting the circles for the ordered system, are the respective exponential fits (of Eq.(\ref{eqn:EL_C})). 
  To compare entanglement lengths between systems with and without disorder, we find that  for example, with $\langle J\rangle/h =0.5$,  $ \xi_{\langle C \rangle}=0.69$ in the disordered case whereas $\xi_C=0.50$ for the ordered one.
The numerical simulations seem to indicate the following: Away from criticality,  the values of concurrence, for all pairs $(i,j)$, are higher in the disordered system as compared to the  corresponding ordered system, signalling order-from- disorder \cite{order-from-disorder-1, order-from-disorder-2, order-from-disorder-3}  as  depicted in Figs.~\ref{fig:decay_conc_xy_glass}(a) and \ref{fig:decay_conc_xy_glass}(d). The roles are reversed as we approach the quantum critical point (see Fig.~\ref{fig:decay_conc_xy_glass}(b)), except when the factorization point is also nearby (see Fig.~\ref{fig:decay_conc_xy_glass}(c)) (cf.~\cite{facto1,facto2}). 
   
 Note that although the panels in Fig.~\ref{fig:decay_conc_xy_glass}  are plotted for system size $N=50$, we have checked that increasing system size does not change the behavior of the entanglements, and hence the entanglement lengths for both disordered and ordered systems.

\subsubsection{Discord length}

We now show that quantum discord length behaves in a qualitatively different way than the entanglement length both in ordered and disordered systems. We will again compare $D_{i,j}$ of the ordered system, govorned by the Hamiltonian $H(J,h)$, with the quenched averaged quantum discord, $\langle D_{i,j}\rangle$, for the $XY$ spin glass system with $\langle J\rangle=J/h$.  To investigate discord length, we divide the entire range of $\langle J\rangle /h$ (for the spin-glass system, which is the same as J/h for the ordered system) into three portions, namely $0<\langle J \rangle /h < \lambda_c$, $\langle J\rangle /h >\lambda_c $  and the neighborhood of \(\lambda_f\), with \(\lambda_c\) and \(\lambda_f\) being the quantum critical and the factorization points of the ordered system respectively. Since we investigate the system by varying $\langle J\rangle/h$, the factorization point lies always in the second region and hence the behavior of discord length in the second region discussed below are excluding the neighborhood of factorization point. The whole discussion will be carried out for $\gamma=0.5$ and hence the factorization point is $\lambda_f=1.1547$. The quantitative feature remain unchanged for other values of $\gamma$.

${\bf{Case~ when~ \langle J \rangle /h < \lambda_c =1}}$: In the ordered system, quantum discord $D_{i,j}$, decays exponentially as in Eq.~(\ref{eqn_EL_D}) \cite{sarandy-DL, amico-DL}. In the  $XY$ spin glass model, quantum discord also decays exponentially, but with a different decay rate. 
As an exemplary case, let us consider $\langle J\rangle /h=0.5$ for which the quenched averaged quantum discord  of the $XY$ spin glass model behaves as
\begin{equation} 
\langle D_{i,j}\rangle=a+ b~e^{-\frac{r}{\xi_{\langle D\rangle}}} \label{eqn:decay_discord_disorder}
\end{equation}
with $a =1.4 \times 10^{-3}$, $b =0.20$,  $\xi_{ \langle D\rangle} =1.36$, while $a = 4.1 \times 10^{-3}$, $b = 0.18$, $\xi_D = 0.56$, for the ordered $XY$ model, implying $\xi_{\langle D\rangle}=2.4 \xi_{ D}$. Therefore, unlike entanglement, we observe significant enhancement of discord length in the disordered system as also depicted in Fig.~\ref{fig:decay_QD_xy_glass}.
The increment of length by introducing randomness in the system  
can therefore only be viewed for quantum correlation measures which are different from entanglement. 
 Moreover, we find that in this region,   $\langle D_{i,j} \rangle > D_{i,j}$, exhibiting thereby an order-from-disorder phenomenon. 

${\bf{Case~ when~\langle J\rangle/ h > \lambda_c = 1}}$: 
Before discussing the disordered system, let us first consider the ordered $XY$ model. 
 In this antiferromagnetic phase, 
 quantum discord saturates to a constant value and hence indicates long-range order in the system even in the thermodynamic limit as also predicted in Refs.~\cite{sarandy-DL, amico-DL}. For example, if one fixes $J/h$ as $1.5$, quantum discord of the $XY$ model without disorder behaves as in Eq.~(\ref{eqn_EL_D}) with $a=0.093, b=0.115$ and $\xi_D=0.80$. Note that we chose $J/h=1.5$ since we are interested in the behavior of discord length which is far from $\lambda_f$, which is $1.1594$ in this case. 
 
 In the $XY$ spin glass model, quenched averaged quantum discord again shows long range order. Specifically, after an initial decay,  it saturates to a constant value. We find  $\xi_{\langle D\rangle}=1.21 > \xi_D=0.80$ for $\langle J \rangle/h=1.5$. However, in this regime, the order from disorder phenomenon is absent since the value of quantum discord of the disordered cases are always lower than that of the corresponding ordered ones (see Fig.~\ref {fig:decay_QD_xy_glass}(d)).  

{\bf{ Neighborhood of  $\lambda_f = \sqrt{1 - \gamma^2}$}}: At the factorization point, quantum discord remains constant for all pairs of \((i, j)\) for the ordered system and hence 
\(\xi_D\) goes to $\infty$, for all non-trivial $b$. In contrast, discord length is finite in the disordered case. Therefore, at the factorization  point, \(\xi_D >  \xi_{\langle D \rangle}\).  The decrement of discord length due to disorder  can also be observed in the vicinity of the factorization point. 
Hence,  enhancement of discord length is  seen in the entire region of  $\langle J \rangle /h$ except  at the neighborhood of the factorization point. 


As a by product, both in the ordered and disordered systems, we can prove that quantum discord does not follow any monogamy relation  \cite{monogamy} for arbitrary $N$, in any of the three regions.  Monogamy of quantum correlation quantitatively states that among three or more parties, if two of them share high amounts of quantum correlation, then they can have only insignificant amounts of quantum correlations with others. Quantitatively, for a given $N$ party state $\rho_{1\dots N}$, a bipartite quantum correlation measure $\cal Q$ is said to be monogamous, with the party 1 as the``nodal observer", if 
\begin{equation}\nonumber
{\cal Q}(\rho_{1:2\dots N}) \geq \sum_{i=2}^{N} { \cal Q} (\rho_{1i}),
\end{equation} 
where ${\cal Q}(\rho_{1:2\dots N})$, and ${\cal Q} (\rho_{1i})$ are respectively the quantum correlation in the $1:\mbox{rest}$ and the $1:i$ bipartition.  \\

 Suppose, if possible, that the quantum discord of the ground state satisfy monogamy. That would imply 
\begin{equation}\nonumber
1\geq {\cal D}(\rho_{1...N}) \geq \sum {\cal D}(\rho_{1i}) \geq ... \geq N {\cal D}(\rho_{1N}).
\end{equation}
The first and the last inequalities are due to the fact that each local system is a qubit and ${\cal D}(\rho_{12}) \geq {\cal D}(\rho_{13}) \geq {\cal D}(\rho_{1N})$ respectively. As argued, ${\cal D}(\rho_{1N})$ can tend to a non-zero constant as $N \rightarrow \infty$, both in the ordered and disordered systems. Therefore, in the thermodynamic limit, $ N {\cal D}(\rho_{1N}) \rightarrow \infty $, giving us a contradiction. Hence there exists some $N$ party quantum state for sufficiently $N$ for which quantum discord does not satisfy the monogamy relation \cite{asu-mon,non-monogamy-QD}. Since $D(\rho_{1N})$ can have a non-zero value for large $N$, it is easy to see that any monogamy-type relations would be violated for quantum discord for those states for sufficiently large $N$. In particular, a similar argument will imply that also the squares of quantum discord also cannot be monogamous for these states with sufficiently large $N$ \cite{discord_square-mon}.
\begin{figure}[h]
\includegraphics[angle=0,width=8.4cm]{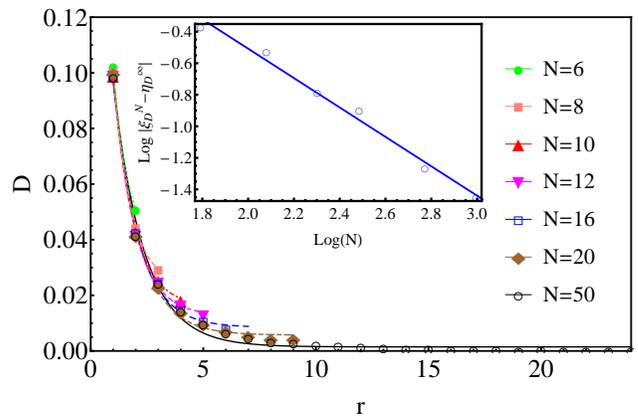} 

\caption{(Color online.) Scalings of discord length for the one-dimensional quantum $XY$ spin glass. We plot the quenched averaged quantum discord on the vertical axis against lattice distance on the horizonal one, for different values of $N$. The logarithms, are with base $e$. In the main figure, the vertical axis is in bits, while the horizontal one is in lattice length units. In the inset, the unit of the horizontal and vertical axes are respectively logarithms of the total numbers of lattice sites and of lattice length.}
\label{fig:QD_xy_glass_scaling}
\end{figure}

It is also interesting to check the behavior of discord length with different $N$. For a fixed $\langle J\rangle/h=0.5$, quenched averaged quantum discord, $\langle D_{i,j}\rangle$, with respect to the lattice distance $r=|i-j|$, for different system sizes is plotted in Fig.~\ref{fig:QD_xy_glass_scaling} for the disordered system.  We calculate ${\xi}^{N}_{\langle D\rangle}$ corresponding to each N. We observe that the behavior of quantum discord freezes for $N\geq50$ and hence we can safely assume that the results obtained for $N=50$ will mimic those of an infinite spin chain. Therefore, we take $ {\xi}_{\langle D \rangle}^{N=50} ={\xi}_{\langle D \rangle}^{\infty}  $. The scaling of the discord length is shown in the inset of Fig.~\ref{fig:QD_xy_glass_scaling} and we find that the discord length scales as $N^{-0.932}$.

\begin{figure}[t]
\includegraphics[angle=0,width=4.2cm]{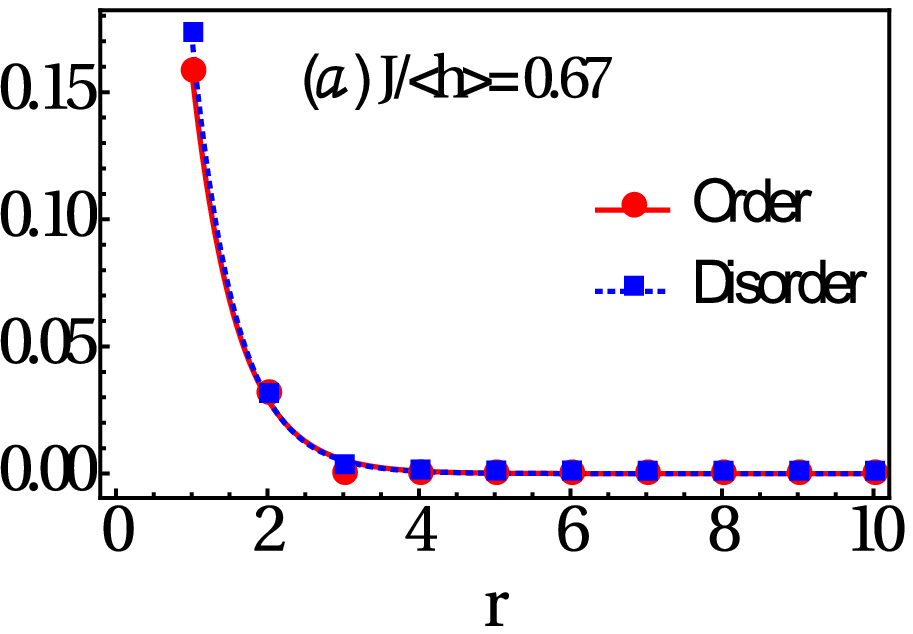} 
\includegraphics[angle=0,width=4.2cm]{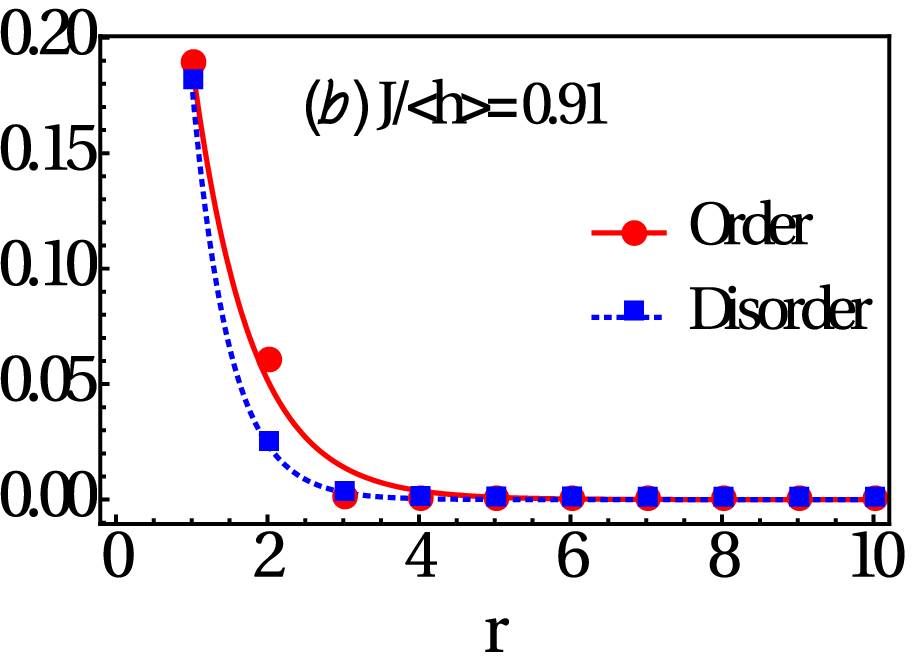} \\
\includegraphics[angle=0,width=4.2cm]{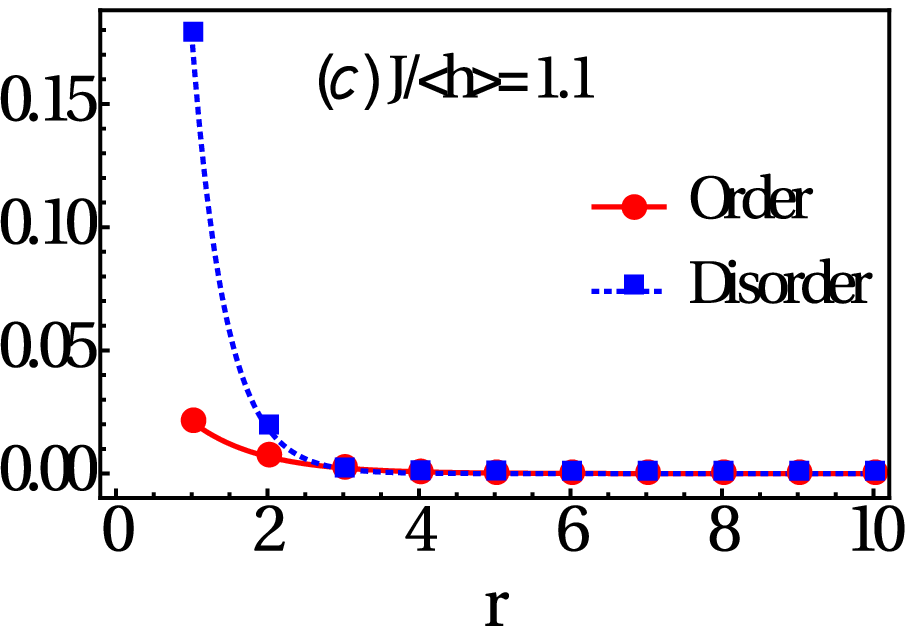} 
\includegraphics[angle=0,width=4.2cm]{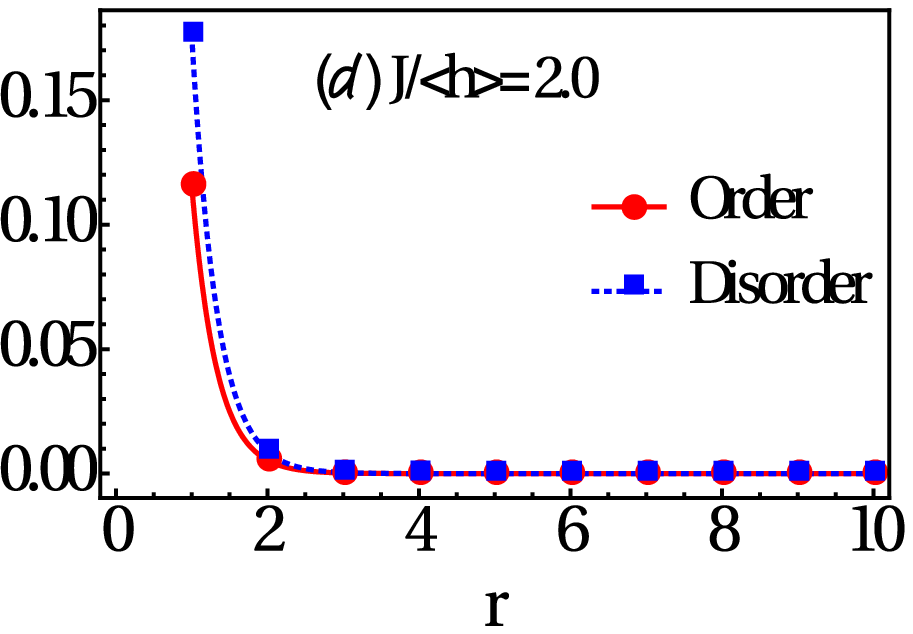} 
\caption{(Color online.) Entanglement length in random field quantum $XY$ spin chain vs. ordered $XY$ chain. All considerations, except for the model considered, and for the fact that the different panels are now for different values of $J/ \langle h\rangle$, remains the same as in Fig.~\ref{fig:decay_conc_xy_glass}.}
\label{fig:decay_conc_xy_field}
\end{figure}
\begin{figure}[t]
\includegraphics[angle=0,width=4.2cm]{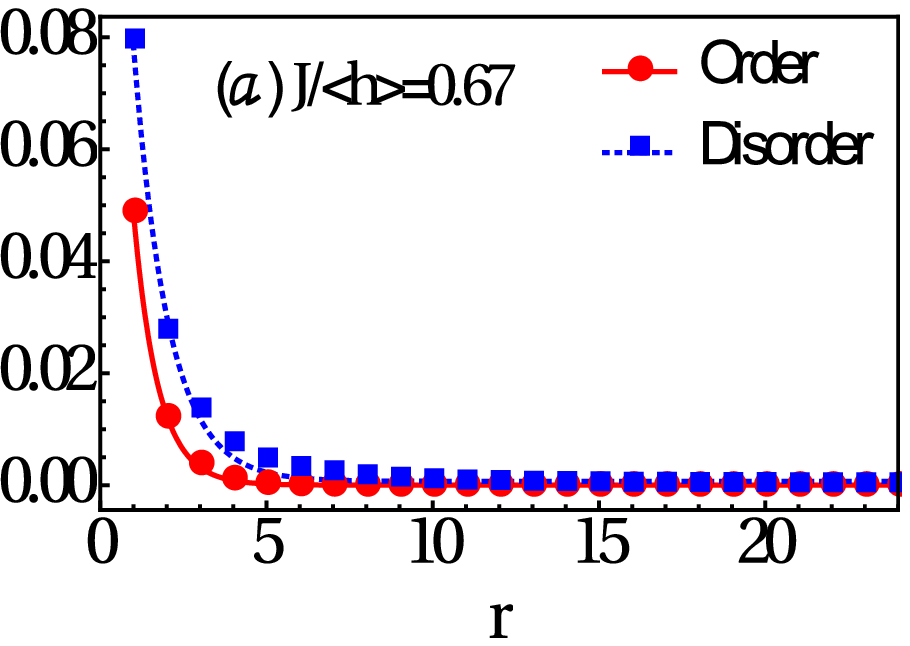} 
\includegraphics[angle=0,width=4.2cm]{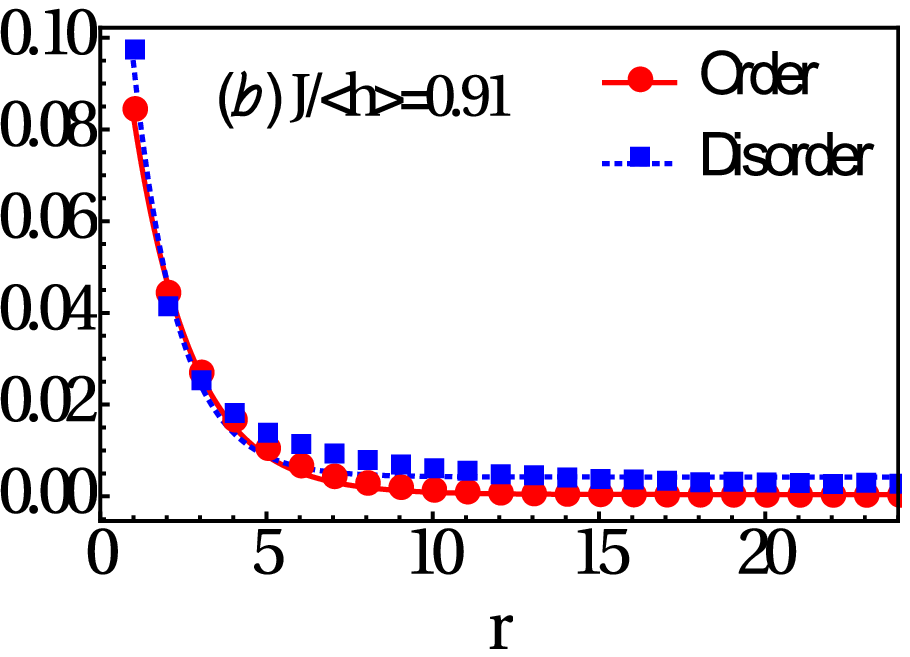} \\
\includegraphics[angle=0,width=4.2cm]{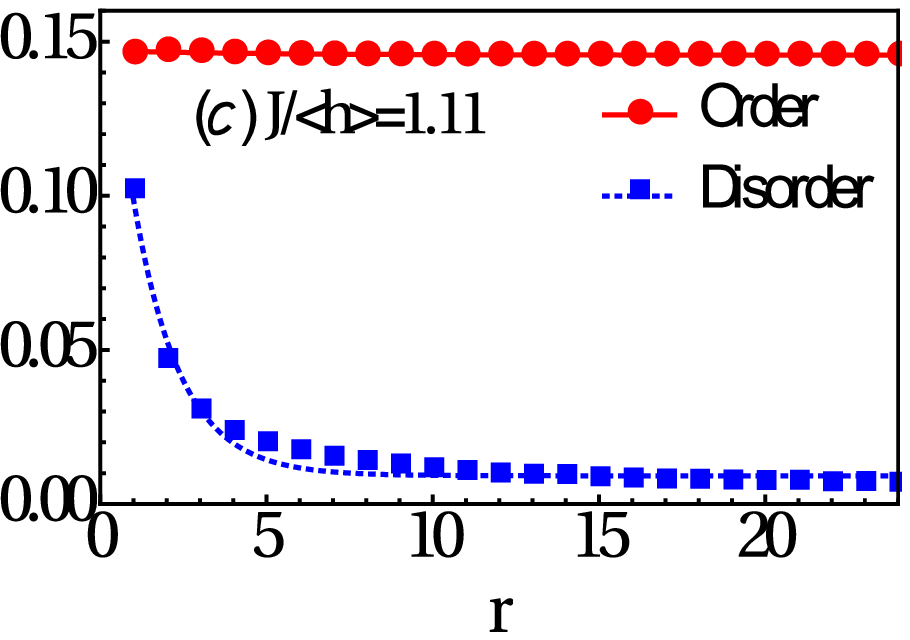} 
\includegraphics[angle=0,width=4.2cm]{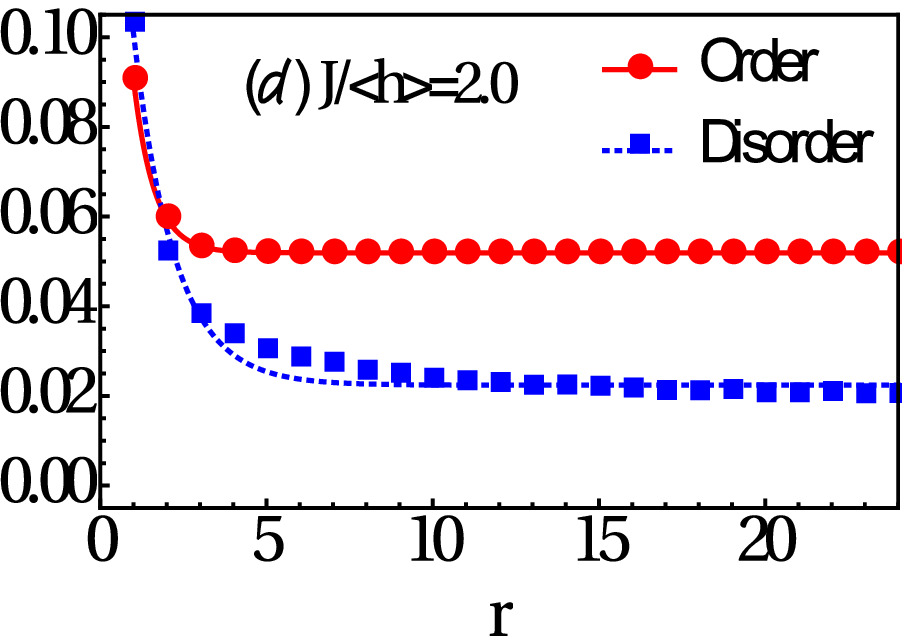} 
\caption{(Color online.) All considerations here are the same as in Fig.~\ref {fig:decay_conc_xy_field}, except that quantum discord is considered instead of concurrence, and that the former is measured in ebits.} 
\label{fig:decay_QD_xy_field}
\end{figure}
\subsection{Random field quantum $XY$ spin chain}
We now consider an $N$-site quantum $XY$ spin chain with uniform nearest neighbor exchange interactions, $J_i=J$, but with field strengths, $h_i$, randomly chosen from $i.i.d.$, the Gaussian probability distributions with mean $\langle h \rangle$ and unit standard deviation. In the corresponding ordered system, $h_i$ assumes a constant value, $\langle h \rangle$, at each site. And again we consider the pure symmetric ground state.

The panels of Figs.~\ref{fig:decay_conc_xy_field} and \ref{fig:decay_QD_xy_field} illustrate the features of concurrence and quantum discord, respectively, for different choices of $\langle h \rangle$, against $r$. We find that the physics remain qualitatively unchanged in the random field $XY$ chain if one compares with the $XY$ spin glasss model. We observe that disorder, in general, does not help in establishing long range entanglement, while it can significantly increase the other quantum correlation length, specifically discord length,  almost everywhere in the parameter space except near the factorization point. As observed for the $XY$ spin glass model,  although discord length, gets enhanced due to disorder, the value of quenched averaged quantum discord decreases in the region $J/\langle h \rangle >1$. On the other hand, the value of entanglement increases in this region, showing a complementarity between the two types of quantum correlation measures. 

\begin{figure}[t]
\includegraphics[angle=0,width=7.8cm]{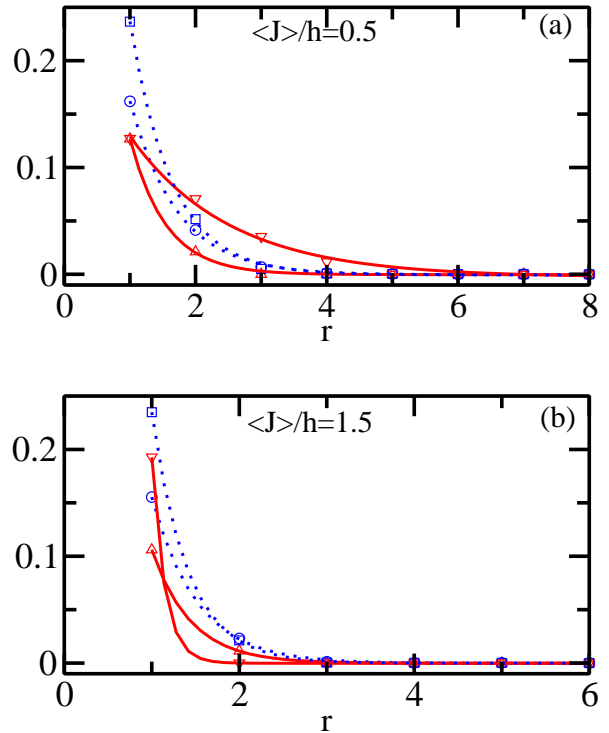}
\caption{(Color online.) The ordered and disordered $XYZ$ models. Scaling of entanglement as a function of distance for the ordered and disordered spin chains with $N=24$ and $\gamma=0.5$ for $ \langle J \rangle/h$ = (a) 0.5 and (b) 1.5. The up and down triangles correspond to the ordered system with $\Delta = 0.1$ and $\Delta = 0.5$, respectively. The circles  and the squares correspond to the disordered system for $\Delta = 0.1$ and $\Delta = 0.5$, respectively. The solid and dotted lines show exponential fits for different cases. For the disordered case, the number of random realizations taken is 8000.}
\label{fig:decay_con}
\end{figure}

\section{ Entanglement length  and Discord length  in $XYZ$ Spin Glass}
\label{sec:XYZ}
In the previous section, we studied quantum correlation length of the quantum $XY$ models. In particular, our results showed that even though the disorder-driven systems are only minimally benefited in terms of the enhancement in concurrence length, a noteworthy endowment occurs in discord length. It is natural to inquire whether the findings are generic in one dimensional systems. Specifically, one can extend the analysis to the disordered quantum $XYZ$ spin glass (see Eq.(~\ref{eqn: XYZ})) in order to find the extent to which the robustness of quantum correlations against the distance between interacting sites are affected further due to the introduction of additional $zz$-interaction, denoted by $\Delta$ (see Eq.~(\ref{eqn: XYZ})). 

As mentioned earlier, the main difficulty in handling a generic one-dimensional system is the absence of an analytical approach akin to the $XY$ model. Therefore, in order to obtain the ground state of the $XYZ$ spin chain, we employ the DMRG technique, which is best suited for studying the spin chains with open boundary conditions in order to achieve high accuracy. However, the drawback is that the measurement of the observables on the fringes would experience boundary effects. In order to investigate the correlations encapsulated between two sites, we consider  the central spin on the $(N/2)^{th}$ site and another site which is positioned at a distance $r$ from the $(N/2)^{th}$  site, but is still far from the boundary. \\

 Let us first concentrate on the entanglement length in the ordered as well as disordered systems.  Fig.~\ref{fig:decay_con} illustrates the behavior of concurrence as a function of distance. We notice that  in the ordered system,  increasing $\Delta$ raises the concurrence length for $\langle J \rangle/h<1$ while there is no notable change in the length, when $\langle J \rangle/h>1$ (see Fig \ref {fig:decay_con}). The situation is true also when disorder in introduced in the system. However, the disordered system fares better than the ordered system in the region $\langle J \rangle/h>1$ compared to that of the $\langle J \rangle/h<1$ for higher values of $\Delta$.
 
Similar to the $XY$ model, quantum discord behaves quite differently than entanglement in the $XYZ$ model with small values of $zz$- interaction, i.e., $\Delta$, essentially mimicking the results obtained for the $XY$ model. Enhancements of discord length are observed both in $\langle J \rangle /h<1$ and $\langle J\rangle/h >1$ regions in the presence of disorder.
 For example, for $\Delta/h=0.1$, and $\langle J\rangle /h=0.5$, $\xi_{\langle D\rangle} = 1.26$ for the disordered system while $\xi_D=0.64$ for the ordered $XYZ$ spin chain. For higher values of $\langle J\rangle /h=1.5$, say,  $\xi_{\langle D \rangle}=1.26$ in the case of disordered system and $\xi_D=1.04$ for the ordered one.\\
 
  However, in the region $\langle J \rangle/h< 1$, the advantage of discord length obtained in the $XY$ spin glass over the corresponding ordered system is faded out with increase of the $zz$-interaction. It also indicates that there exists an interplay between the randomness in the coupling strength and the non-random $zz$-interaction. In particular, for fixed $\langle J\rangle/h$, with the increase of $\Delta$, the discord length decreases in the disordered system. This is illustarted in Table \ref{table:nonlin}, where we consider a chain of 24 spins.
 \begin{table}[ht]
\centering
\begin{tabular}{|c| c | c|} 
\hline\hline 
$ $   &  $ \Delta/h=0.1$  &  $\Delta/h=0.5$ \\ 
\hline\hline 
 $\langle J \rangle /h=0.5$               &      $\xi_D = 0.64$                     &     $\xi_D=4.05$                    \\ 
                &      $\xi_{\langle D\rangle} = 1.26$                     &     $\xi_{\langle D\rangle}=0.86$                    \\ 
\hline
  $\langle J \rangle /h=1.5$              &     $\xi_D = 1.04 $                  &     $\xi_D =   0.68  $               \\  
              &     $\xi_{\langle D\rangle} = 1.26 $                  &    $\xi_{\langle D\rangle} =   0.73 $               \\  
\hline
\end{tabular}
\caption{Comparison of discord length for both the ordered and disordered systems in quantum $XYZ$ model  for different values of $\langle J \rangle /h$ and $\Delta/h$ with $N=24$. }
\label{table:nonlin} 
\end{table}
 

 \begin{figure}[t]
\includegraphics[angle=0,width=7.8cm]{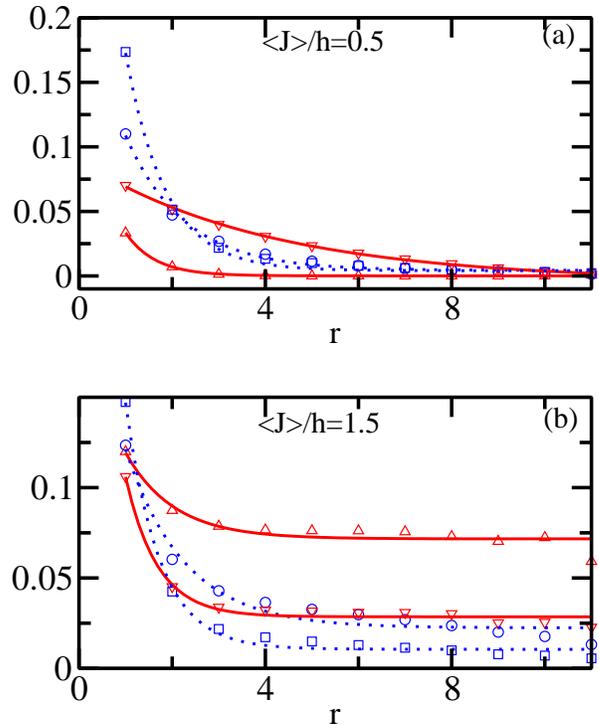}
\caption{(Color online.)  All considerations here are the same as in Fig.~\ref{fig:decay_con}, except that quantum discord is considered instead of concurrence, and that the vertical axes are measured in bits.}
\label{fig::decay_dis}
\end{figure}
\section{Conclusion}
\label{sec:conclusion}

In summary, we concentrate mainly on the effects of disorder in quantum correlation lengths of spin chains. The paradigmatic models that we consider are $XY$ spin glass model in which coupling strengths are chosen randomly, the $XY$ spin chain with random field, and the $XYZ$ model with random $xx$- and $yy$-couplings. Specifically, we compare quantum correlation lengths, namely entanglement and quantum discord lengths, of the disordered models with the corresponding ordered systems. \\

We find that entanglement length neither shows significant reduction nor increase with the introduction of disorder in the system. In sharp contrast, the discord length is significantly higher in the $XY$ disordered model in comparision to the corresponding ordered ones. The features remain unaltered in the $XYZ$ model for low values of the $zz$-interactions. Higher values of the $zz$-interaction, however, interfares destructively with the disordered couplings, to supress the disorder-induced enhancement in discord length. As a consequence of the fact that discord remains finite between arbitrary two sites, we prove that quantum discord cannot satisfy a large class of monogamy relations for the ground states of the ordered and disordered spin models for sufficiently large systems. 
\acknowledgments

RP acknowledges support from the Department of Science and Technology, Government of India, in the form of an INSPIRE faculty scheme at HRI. We acknowledge computations performed at the cluster computing facility in HRI. This work has been developed by using the DMRG code released within the ``Powder with Power" project (\url{http://www.qti.sns.it}).



\begin{thebibliography}{10}

\bibitem{Horodecki}
R. Horodecki, P. Horodecki, M. Horodecki, and K. Horodecki, Rev. Mod. Phys. {\bf 81}, 865 (2009).
\bibitem{AditiUjjwal}
A. Sen(De), U. Sen, Physics News,  {\bf 40}, 17 (2010) (arXiv: 1105.2412). 


\bibitem{densecode}
C. H. Bennett and S. J. Wiesner, Phys. Rev. Lett. {\bf 69}, 2881 (1992); K. Mattle, H. Weinfurter, P. G. Kwiat, and A. Zeilinger, Phys. Rev. Lett. {\bf 76}, 4656 (1996).

\bibitem{teleportation}
C. H. Bennett, G. Brassard, C. Cr{\'e}peau, R. Jozsa, A. Peres, and W. K. Wootters, Phys. Rev. Lett. {\bf 70}, 1895 (1993); D. Bouwmeester, J. -W. Pan, K. Mattle, M. Eibl, H. Weinfurter, and A. Zeilinger, Nature {\bf 390}, 575 (1997).

\bibitem{BB84}
A. K. Ekert, Phys. Rev. Lett. {\bf 67}, 661 (1991); N. Gisin, G. Ribordy, W. Tittel, and H. Zbinden, Rev. Mod. Phys. { \bf74}, 145 (2002). 

\bibitem{briegel}

R. Raussendorf and H. J. Briegel, Phys. Rev. Lett. { \bf 86},
5188 (2001); F. Meier, J. Levy, and D. Loss, ibid.{\bf  90},
047901 (2003); R. Raussendorf, D. E. Browne, and H. J.
Briegel, Phys. Rev. A {\bf 68}, 022312 (2003); M. A. Nielsen,
Phys. Rev. Lett. { \bf 93}, 040503 (2004); P. Walther, K. J.
Resch, T. Rudolph, E. Schenck, H. Weinfurter, V. Vedral,
M. Aspelmeyer, and A. Zeilinger, Nature {\bf 434}, 169
(2005); M. A. Nielsen, Rep. Math. Phys.{ \bf 57}, 147 (2006);
H. J. Briegel, D. E. Browne, W. D{\" u}r, R. Raussendorf,
and M. V. den Nest, Nat. Phys. { \bf 5}, 19 (2009).







\bibitem{manybody-entanglement}
 M. Lewenstein, A. Sanpera, V. Ahufinger, B. Damski, A. Sen(De), and U. Sen, Adv. Phys. {\bf 56}, 243 (2007); L. Amico, R. Fazio, A. Osterloh, and V. Vedral, Rev. Mod. Phys. {\bf 80}, 517 (2008).
 
 \bibitem{bose}
 S. Bose, Phys. Rev. Lett. {\bf 91}, 207901 (2003), and referencs there to.
 
 \bibitem{quant-compu-1}
R. P. Feynman, 1982 Int. J. theor. Phys. {\bf 21}, 467; A. M. Turing, 1936 Proc. Lond. Math. Soc. Ser. {\bf 2}, 442, 230; M. Nielsen and I. Chuang, {\emph Quantum computation
and quantum information} (Cambridge University Press, Cambridge, 2000).





\bibitem{cold-atom}
I. Bloch, J. Phys. B: At. Mol. Opt. Phys. {\bf 38}, S629 (2005); P. Treutlein, T. Steinmetz, Y. Colombe, B. Lev, P. Hommelhoff, J. Reichel, M. Greiner, O. Mandel, A. Widera, T. Rom, I. Bloch, and T. W. H{\"a}nsch, Fortschr. Phys. {\bf 54}, 702 (2006).

\bibitem{ions}
D. Leibfried, R. Blatt, C. Monroe, and D. Wineland, Rev. Mod. Phys. {\bf 75}, 281 (2003); H. H{\"a}ffner, C. Roos, and R.Blatt, Phys. Rep. {\bf 469} 155 (2008).

\bibitem{NMR}
L. M. K. Vandersypen and I. L. Chuang. Rev. Mod. Phys. {\bf76}, 1037.


\bibitem{heisenberg-material}
M. D. Johannes, J. Richter, S.-L. Drechsler, and H. Rosner, Phys. Rev. B {\bf 74}, 174435 (2006). 

\bibitem{order-from-disorder-1}
G. Misguich and C. Lhuillier, Frustrated Spin Systems, edited by H. T. Diep (World-Scientific, Singapore, 2005).

\bibitem{network}
M. Zukowski, A. Zeilinger, M. A. Horne, and A. K. Ekert, Phys. Rev. Lett., {\bf 71}, 4287 (1993); H. J. Briegel, W. D{\" u}r, J. I. Cirac, and P. Zoller, Phys. Rev. Lett., {\bf 81}, 5932 (1998); W. D{\" u}r, H. J. Briegel, J. I. Cirac, and P. Zoller, Phys. Rev. A {\bf 59}, 169 (1999); N. Sangouard, C. Simon, H. de Riedmatten, and N. Gisin, Rev. Mod. Phys. {\bf 83}, 33 (2011).





\bibitem{lieb-robin}

E. H. Lieb and D. W. Robinson, Commun. Math. Phys. {\bf{28}}, 251 (1972).

\bibitem{eisert_tensor}

J. Eisert,  arXiv: 1308.3318.
\bibitem{hasting}

M. B. Hastings, T. Koma,  Commun. Math. Phys. { \bf 265}, 781 (2006) .

\bibitem{phase-transition-ent}
A. Osterloh, L. Amico, G. Falci, and R. Fazio, Nature {\bf 416},  608 (2002).

\bibitem{hisen_length}
B.-Q. Jin and V. E. Korepin, Phys. Rev. A {\bf 69}, 062314  2004.


 
 
 
 



\bibitem{ref-discord}
 L. Henderson and V. Vedral, J. Phys. A: Math. Gen. {\bf 34}, 6899 (2001);  H. Ollivier and W. H. Zurek, Phys. Rev. Lett. {\bf 88}, 017901 (2001).



\bibitem{ref-wd}
J. Oppenheim, M. Horodecki, P. Horodecki, and R. Horodecki, Phys. Rev. Lett. {\bf 89}, 180402 (2002); M. Horodecki, K. Horodecki, P. Horodecki, R. Horodecki, J. Oppenheim, A. Sen(De), and U. Sen,{ \it ibid}. {\bf 90}, 100402 (2003); I. Devetak, Phys. Rev. A {\bf 71}, 062303 (2005); M. Horodecki, P. Horodecki, R. Horodecki, J. Oppenheim, A. Sen(De), U. Sen, and B. Synak-Radtke, Phys. Rev. A {\bf 71}, 062307 (2005).

\bibitem{negative-eff-defect}
K. Modi, A. Brodutch, H. Cable, T. Paterek, and V. Vedral, Rev. Mod. Phys. {\bf 84}, 1655 (2012).

\bibitem{phase-transition-discord1}
R. Dillenschneider, Phys. Rev. B {\bf 78}, 224413 (2008).

\bibitem{phase-transition-discord}
 T. Werlang, C. Trippe, G. A. P. Ribeiro, and G. Rigolin, Phys. Rev. Lett. {\bf 105}, 095702 (2010); T. Werlang, G. A. P. Ribeiro, and G. Rigolin, Phys. Rev. A {\bf 83}, 062334 (2011).

\bibitem{sarandy-DL} M. S. Sarandy, Phys. Rev. A \textbf{80}, 022108 (2009); J. Maziero, H. C. Guzman, L. C. Celeri, M. S. Sarandy, and R. M. Serra, Phys. Rev. A {\bf 82}, 012106 (2010); J. Maziero, L.C. Céleria, R.M. Serra and M.S. Sarandy, Phys. Lett. A \textbf{376}, 1540 (2012).

\bibitem{amico-DL}
M. S. Sarandy, T. R. De Oliveira and L. Amico, Int. J. Mod. Phys. B \textbf{27}, 1345030 (2013).




\bibitem{disorder-expt}
L. Fallani, J. E. Lye, V. Guarrera, C. Fort, and M. Inguscio, Phys. Rev. Lett. {\bf 98}, 130404 (2007); G. Roati, C. D'Errico, L. Fallani, M. Fattori, C. Fort, M. Zaccanti, G. Modugno, M. Modugno, and M. Inguscio, Nature {\bf 453}, 895 (2008); J. Billy, V. Josse, Z. Zuo, A. Bernard, B. Hambrecht, P. Lugan, D. Cl´ement, L. Sanchez-Palencia, P. Bouyer, and A. Aspect, {\it ibid}. {\bf 453}, 891 (2008); R. Yu, L. Yin, N. S. Sullivan, J. S. Xia, C. Huan, A. Paduan-Filho, N. F. Oliveira Jr., S. Haas, A. Steppke, C. F. Miclea, F. Weickert, R. Movshovich, E.-D. Mun, V. S. Zapf, and T. Roscilde, {\it ibid}. {\bf 489}, 379 (2012); D. H¨vonen,
S. Zhao, M. M˚ansson, T. Yankova, E. Ressouche, C. Niedermayer, M. Laver, S. N. Gvasaliya, and A. Zheludev,
Phys. Rev. B { \bf85}, 100410(R) (2012); S. Krinner, D. Stadler, J. Meineke, J.-P. Brantut, and T. Esslinger, arXiv: 1311.5174 [quant-ph]; K. R. A. Hazzard, B. Gadway, M. Foss-Feig, B. Yan, S. A. Moses, J. P. Covey, N.
Y. Yao, M. D. Lukin, J. Ye, D. S. Jin, and A. M. Rey, arXiv: 1402.2354 [quant-ph], and references therein.


\bibitem{order-from-disorder-2}
A. Aharony, Phys. Rev. B {\bf 18}, 3328 (1978); J. Villain, R. Bidaux, J.-P. Carton, and R. Conte, J. Physique {\bf 41}, 1263 (1980); B. J. Minchau and R. A. Pelcovits, Phys. Rev. B {\bf 32}, 3081 (1985); C. L. Henley, Phys. Rev. Lett. {\bf 62}, 2056 (1989); A. Moreo, E. Dagotto, T. Jolicoeur, and J. Riera, Phys. Rev. B {\bf 42}, 6283 (1990); D. E. Feldman, J. Phys. A {\bf 31}, L177 (1998); G. E. Volovik, JETP Lett.
{\bf 84}, 455 (2006); D. A. Abanin, P. A. Lee, and L. S. Levitov, Phys. Rev. Lett. \textbf{98}, 156801 (2007); L. Adamska, M. B. Silva Neto, and C. Morais Smith, Phys. Rev. B \textbf{75}, 134507 (2007); A. Niederberger, T. Schulte, J. Wehr, M. Lewenstein, L. Sanchez-Palencia, and K. Sacha, Phys. Rev. Lett. \textbf{100}, 030403 (2008); A. Niederberger, J. Wehr, M. Lewenstein, and K. Sacha, Europhys. Letts. \textbf{86}, 26004 (2009); A. Niederberger, M. M. Rams, J. Dziarmaga, F. M. Cucchietti, J. Wehr, and M. Lewenstein, Phys. Rev. A \textbf{82}, 013630 (2010); D. I. Tsomokos, T. J. Osborne, and C. Castelnovo, Phys. Rev. B \textbf{83}, 075124 (2011); M. S. Foster, H.-Y. Xie, and Y.-Z. Chou, {\it ibid}. \textbf{89}, 155140 (2014);
P. Villa Mart´ın, J. A. Bonachela, and M. A. Mu˜noz, Phys. Rev. E \textbf{89}, 012145 (2014), and references therein.

\bibitem{order-from-disorder-3}
L. F. Santos, G. Rigolin, and C. O. Escobar, Phys. Rev. A \textbf{69}, 042304 (2004); C. Mej´ıa-Monasterio, G. Benenti, G. G. Carlo, and G. Casati, {\it ibid}. \textbf{71}, 062324 (2005); A. Lakshminarayan
and V. Subrahmanyam, {\it ibid}. \textbf{71}, 062334 (2005); R. L´opez-Sandoval and M. E. Garcia, Phys. Rev.
B \textbf{74}, 174204 (2006); J. Karthik, A. Sharma, and A. Lakshminarayan, Phys. Rev. A \textbf{75}, 022304 (2007); W. G. Brown, L. F. Santos, D. J. Sterling, and L. Viola, Phys. Rev. E \textbf{77}, 021106 (2008); F. Dukesz, M. Zilbergerts, and L. F. Santos, New J. Phys. \textbf{11}, 043026 (2009); J. Hide, W. Son, and V. Vedral, Phys. Rev. Lett. \textbf{102}, 100503 (2009); K. Fujii and K. Yamamoto, Phys. Rev. A \textbf{82},
042109 (2010); R. Prabhu, S. Pradhan, A. Sen(De), and U. Sen, Phys. Rev. A \textbf{84}, 042334 (2011); U. Mishra, D. Rakshit, R. Prabhu, A. Sen(De), and U. Sen arXiv:1408.0179 and references therein.




\bibitem{LSM} E. Lieb, T. Schultz, and D. Mattis, Ann. Phys. \textbf{16}, 407 (1961).

\bibitem{barouch1} E. Barouch, B. McCoy, and M. Dresden, Phys. Rev. A \textbf{2}, 1075 (1970).

\bibitem{barouch2} E. Barouch and B. McCoy, Phys. Rev. A \textbf{3}, 786 (1971).


\bibitem{dmrg} S. R. White, Phys. Rev. Lett. {\bf 69}, 2863 (1992); Phys. Rev. B {\bf 48}, 10345 (1993); U. Schollw\"ock, Rev. Mod. Phys. {\bf 77}, 259 (2005).


\bibitem{wooters} S. Hill and W. K. Wootters, Phys. Rev. Lett. {\bf 78}, 5022 (1997); W. K. Wootters, {\it ibid}. {\bf 80}, 2245 (1998).

\bibitem{mutual1}
W. H. Zurek, in Quantum Optics, Experimental Gravitation and
Measurement Theory, edited by P. Meystre and M. O. Scully
(Plenum, New York, 1983); S. M. Barnett and S. J. D. Phoenix,
Phys. Rev. A {\bf 40}, 2404 (1989).

\bibitem{mutual2}
 N. J. Cerf and C. Adami, Phys. Rev. Lett. {\bf 79}, 5194 (1997).
 
 \bibitem{mutual3}
B. Schumacher and M. A. Nielsen, Phys. Rev. A { \bf 54}, 2629 (1996);
B. Groisman, S. Popescu, and A. Winter, {\it ibid}. { \bf 72}, 032317
(2005).



\bibitem{x-states} T. Yu and J. H. Eberly, arXiv:quant-ph/0503089.

\bibitem{Ali-et-all}
M. Ali, A. R. P. Rau, and G. Alber, Phys. Rev. A \textbf{81}, 042105 (2010). 

\bibitem{Wang}
 X.-M. Lu, J. Ma, Z. Xi, and X. Wang, Phys. Rev. A \textbf{83}, 012327 (2011).

\bibitem{Chen} Q. Chen, C. Zhang, S. Yu, X. X. Yi, and C. H. Oh, Phys. Rev. A
\textbf{84}, 042313 (2011).



\bibitem{Huang-Disc-num} Y. Huang, Phys. Rev. A \textbf{88}, 014302 (2013).



\bibitem{LE}
 F. Verstraete, M. Popp, and J. I. Cirac, Phys. Rev. Lett. {\bf 92}, 027901 (2004); F. Verstraete, M. A. Martín-Delgado, and J. I. Cirac, {\it ibid}. {\bf 92}, 087201 (2004).

\bibitem{dorit}
 
D. Aharonov, Phys. Rev. A { \bf 62}, 062311.
 
 \bibitem{popsescu_rol}
 
 S. Popescu, D. Rohrlich, Phys. Lett. A { \bf166}, 293 (1992).
 
 \bibitem{asd_sen_marek}
 
 A. Sen(De), U. Sen, M. Wiesniak, D. Kaszlikowski, M. Zukowski. Phys. Rev. A { \bf 68}, 062306.
 
 
\bibitem{zanardi} S. Garnerone, N.T. Jaconson, S. Hass, P. Zanardi, Phys.
Rev. Lett. {\bf 102}, 057205 (2009); N.T. Jacobson, S. Garnerone, S. Hass, P. Zanardi, Phys. Rev. B {\bf 79}, 184427 (2009).

\bibitem{group-disorder} D. Sadhukhan, S. Singha Roy,  D. Rakshit, A. Sen(De), and U. Sen, New J. Phys. (in print) (arXiv:1406.7239); D. Sadhukhan, R. Prabhu,  A. Sen(De), and U. Sen, arXiv:1412.8385.




\bibitem{facto1}

J. Kurmann, H. Thomas, and G. Muller, Physica A { \bf112}, 235 (1982); G. M\"uller, R.E. Shrock, Phys. Rev. B {\bf 32}, 5845 (1985).

\bibitem{facto2}

J. Eakins and G. Jaroszkiewicz, J. Phys. A: Math. Gen. {\bf 36}, 517 (2003); T. Roscilde, P. Verrucchi, A. Fubini, S. Haas, and V. Tognetti, Phys. Rev. Lett. {\bf 93}, 167203 (2004), {\it ibid}. {\bf 94}, 147208 (2005); S. Dusuel and J. Vidal, Phys. Rev. B {\bf 71}, 224420 (2005); L. Amico, F. Baroni, A. Fubini, D. Patanè, V. Tognetti, and Paola Verrucchi, Phys. Rev. A {\bf 74}, 022322 (2006); F. Baroni, J. Phys. A {\bf 40}, 9845 (2007); F. Baroni, A. Fubini, V. Tognetti and P. Verrucchi, {\it ibid.} {\bf 40}, 9845 (2007);
S. M. Giampaolo, G. Adesso, and F. Illuminati. Phys. Rev. Lett. {\bf 100}, 197201 (2008); B. Cakmak, G. Karpat and F. F. Fanchini, arXiv:1502.02306 (2015). 















\bibitem{monogamy} C. H. Bennett, H. J. Bernstein, S. Popescu, and B. Schumacher, Phys. Rev. A { \bf 53}, 2046 (1996); V. Coffman, J. Kundu, and W. K. Wootters,{ \it ibid}. \textbf{61}, 052306 (2000); T. Osborne and F. Verstraete, Phys. Rev. Lett. {\bf 96}, 220503 (2006); G. Adesso, A. Serafini, and F. Illuminati, Phys. Rev. A {\bf 73} , 032345 (2006); Y. -C. Ou and H. Fan,{\it ibid}. { \bf 75}, 062308 (2007); T. Hiroshima, G. Adesso, and F. Illuminati, Phys. Rev. Lett. {\bf 98}, 050503 (2007); M. Seevinck, Phys. Rev. A {\bf 76}, 012106 (2007); S. Lee and J. Park, {\it ibid}. {\bf 79}, 054309 (2009); A. Kay, D. Kaszlikowski, and R.Ramanathan, Phys. Rev. Lett. {\bf 103}, 050501 (2009); M.Hayashi and L. Chen, Phys. Rev. A {\bf 84}, 012325 (2011), and references therein.

\bibitem{asu-mon}A. Kumar, R. Prabhu, A. Sen(De), and U. Sen  Phys. Rev. A { \bf 91}, 012341 (2015). 

\bibitem{non-monogamy-QD} M. Allegra, P. Giorda, and A. Montorsi, Phys. Rev. B \textbf{84}, 245133 (2011).


\bibitem{discord_square-mon}
Y. Bai, N. Zhang, M. Ye, and Z. D. Wang. Phys. Rev. A {\bf88}, 012123 (2013).





\end{thebibliography}
\end{document}